# Spatial offsets between interstellar bone-like filaments, radio masers, and cold diffuse CO gas, in the Scutum spiral arm

Jacques P. Vallée

National Research Council of Canada, Herzberg Astronomy & Astrophysics Research Centre, 5071 West Saanich Road, Victoria, B.C., Canada V9E 2E7   jacques.p.vallee@gmail.com

**Abstract.**
The advent of more precise measurements of distance estimates for some objects (bayesian estimates for filaments, trigonometric estimates for masers) permits a better comparison of their relative locations, and a comparison with a most recent spiral arm model fitted to the diffuse CO 1-0 gas. Our results support the idea that some 'bone-like' filaments, greater than 10 pc (labeled elsewhere as 'bones', 'mst', 'herschel') and smaller than 100 pc, are observed near the potential minimum of the Scutum spiral arm (bone-like filaments are offset outward from masers by about 200 pc).

### 1. Introduction

In the interstellar medium, there are many objects called filaments in the literature, and we are interested in where they are located with respect to an arm. A recent short review of interstellar filaments was made elsewhere (Vallée 2017c – section 4.2).

Filament types. Not all filaments are alike. Thus some authors are now assigning different 'types' of filaments, e.g., 'bone', 'mst', herschel', 'gmf' (see column 5 in table 1 of Zucker et al 2018). Some acronyms are employed in the literature: thus 'bones' refers to high-density filamentary ridges (Goodman et al 2014; Zucker et al 2015), while 'mst' is spelled 'minimum spanning tree' (see Wang et al 2016), 'herschel' stands for 'large scale Herschel filament' (see Wang et al 2015), 'gmf' means 'giant molecular filament' (see Ragan et al 2014; Abreu-Vicente et al 2016).

 Extraction techniques. Filaments are observed in different ways. Some definitions employed a velocity-continuous criterion, as well as a position-position-continuous criterion. Some used 'bye-eye' or 'automated' searches, 'blind' or 'spiral-tracing' filaments, arm association within 10 km/s or less. Some require a proximity to the Galactic plane.

Disparities in their extraction techniques and in their physical properties may point to different origins for filaments. To compute physical properties, a good distance estimate is required. Zucker et al (2018 – their fig.4) calculated the

Bayesian distance to each filament, taking account of trigonometric parallax, modeling, velocity, latitude distance from the disk, and nearest arm proximity.

**Some are short:** all 413 filaments studied by Mattern et al (2018), seen in $^{13}CO$ 2-1 and $C^{18}O$ 2-1, have a length between 0.02 pc to 1.8 pc (their table 6). Their velocity distances place a large fraction of the filaments near the Scutum-Crux-Centaurus spiral arm (their Fig. 9).

**Some are long:** Zhang et al (2019) uniformly analysed 57 giant filaments exceeding 10 pc, looking for their star-forming content. The filaments of Zhang et al (2019) have a median length of 67 pc, and a solar mass of $15 \times 10^4$, similar to the 4$^{th}$ type (gmf). Also, a standardized analysis of 45 giant filaments exceeding 10 pc, looking for their physical properties, was made by Zucker et al (2018).

**Some are longer.** In comparison, the 4$^{th}$ filament type, gmf, has much larger median values (length of 84 pc, with of 12 pc, solar mass of $18 \times 10^4$, column density of $5 \times 10^{21}$ cm$^{-2}$). Zucker et al (2018) noted that the 'gmf' filaments have a lower column density fraction (their fig. 8), and a higher length, width and linear mass (their fig.10), and a lower aspect ratio (their fig. 16) than the 3 other types (bone, mst, herschel). The 4$^{th}$ type (gmf) does not have 'bone-like' properties (too long, too large, too massive).

In Zucker et al (2018), the first three types have similar physical properties: median length (36 pc for bones, 27 pc for mst, 50 pc for herschel), median width (2 pc for bones, 4 pc for mst, 2 pc for herschel), median solar mass ($1.5 \times 10^4$ for bones, $2.1 \times 10^4$ for mst, $2.6 \times 10^4$ for herschel), median column density ($9 \times 10^{21}$ cm$^{-2}$ for bones, $12 \times 10^{21}$ cm$^{-2}$ for mst, $5 \times 10^{21}$ cm$^{-2}$ for Herschel). Zucker et al (2018) associated their own 'Bone' filaments with the mst 'bones' (their fig. 6, 7, 12 and section 4.2) and with herschel 'bones' (their fig. 16 and Section 4.3).

While the 'gmf' type seems a broad term relating to a large scale feature, it is desirable to look for sub-sets (like 'bones') that can be better observed and parameterized (with a better distance available) – see Zucker et al (2018).

What is the criteria for a filament to be 'bone-like'? A narrow filament category can be defined as 'bone-like', which would include the three earlier types of 'bone', 'mst', and 'herschel' filaments. This 'bone-like' criteria may have a high column density filamentary ridge ($>5 \times 10^{21}$ cm$^{-2}$ – see above), with a physical size not too short (>10 pc) but not too long (<100 pc). A part of a long 'gmf' could be termed a bone-like filament; along the long ridge of the gmf its column density may vary greatly; the whole gmf may be more massive (too long, too large).

Thus this 'bone-like' category of filaments (bone, mst, herschel) has a range of length from 10 pc to 100 pc, a width from 1 pc to 6 pc, and a solar mass from $0.4 \times 10^4$ to $7 \times 10^4$ (all data from Zucker et al 2018). In galactic altitude, they are all located within 60 pc of the galactic mid-plane. These bone-like filaments thus exclude all shorter filaments in Mattern at al (2018).

In this paper, we focus on 'bone-like' filaments, not too short (>10 pc) and not too long (<100 pc), as defined above. In Section 2, we use the published Bayesian distance for each bone-like filament, as well as the trigonometric distance for each known maser, and a recently published 4-arm model for the Galaxy based on a fit to the arm tangents using the peak of the diffuse CO gas. A majority (25/45) of such filaments with a Bayesian distance estimate are associated to the 'Scutum' arm by Zucker et al (2018), so we selected the Scutum-Crux-Centaurus arm for a closer examination. We also complement this top-view with a longitude velocity view for the same recent 4-arm model.

Where could the 'bone-like' filament be located, along their orbit: in or out of a spiral arm? In a density-wave galactic model, the masers are expected to have formed in or close to the shocked 'dust lane', which is near the location of the 'density-maximum', while the cold diffuse CO gas is farther out along the gas orbit (Roberts 1975– his Fig. 2). The typical spiral arm cross-section shows a dust lane in the inner arm side, followed closely by a lane of masers indicative of very young star formation. Farther out from the dust/maser lanes, young stars and HII regions ensue, and then the peak intensity of the cold diffuse CO 1-0 gas seen when scanning a telescope beam in galactic longitudes, to be followed by many older stars (Vallée 2014 – his Fig.2).

We will compare these structures with the position of the recent 4-arm models of Vallée (2017a; 2017b), as it is rigorously fitted to the catalog of arm tangents seen in CO 1-0 gas (obtained by scanning in galactic longitudes, and listed in Vallée 2016). The observed CO 1-0 is far more abundant to observe than some other tracers such as masers. This model employs a distance of the Sun to the Galactic Center of 8.0 kpc (Vallée, 2017d).

In Section 3, we delve into the possible origins of the bone-like filaments, a subsection of all filaments. In Section 4 we discuss the bone-like filaments and published theories, and we conclude in Section 5.

## 2. Where are bone-like filaments and spiral arms (top view and velocities)

In this paper, we restrict ourselves to the first three types (bone, mst, herschel) in the paper of Zucker et al (2018).

In **Table 1** here, we assembled all filaments in Zucker et al (2018) in or near the Scutum-Crux-Centaurus arm, as well as all trigonometric masers in Reid et al (2014; 2009) in or near the Scutum-Crux-Centaurus arm.

The filament distances are from Zucker at al (2018), employing the latest Bayesian procedure; their Fig. 4 shows how to combine probabilistically the kinematic distance (1 kpc error), the proximity of an arm (0.5 kpc error), the proximity of a gmc with a parallax measure (0.3 kpc error), galactic latitude, and a prior probability that it lies at a far distance (0.01). The final Bayesian result for the

Snake filament had a distance error near 0.3 kpc. If all filaments near the Scutum arm are sufficiently similar, then the mean distance could be obtained with a much smaller mean error (the standard deviation of the mean is smaller than that by the inverse square root of the number of filaments employed), or roughly 3 or more times smaller for 10 filaments.

The Bayesian model employed for distance determination by Zucker et al (2018) relies on the spiral arm model of Reid et al (2016). The choice of a nearby arm assignment versus a distant arm assignment is not likely to change with another recent 4-arm model. The Reid model is discussed below. However, one could worry that the Bayesian distance estimator takes into account the proximity to a nearby spiral arm for determining the most likely distance, and this factor may introduce some bias in assuming that these sources may follow the spiral structure and in adjusting the distances to put them into the spiral structure – we will test this possibility later (Section 2.3).

2.1 The 4-arm model

We then employed the latest spiral arm model, as fitted to the diffuse CO 1-0 arm tangents (listed in Vallée 2016), and as computed and published (Vallée 2017a; 2017b). This diffuse CO-fitted arm would correspond to the 'potential minimum' of a density-wave (e.g., Fig. 2 in Roberts 1975). We emphasize that this 4-arm model was fitted to the observed tangent (from the sun to each arm) of the diffuse CO 1-0 gas, as observed along the galactic plane to be peaking in intensity at specific galactic longitudes (see Tables 3 and 5 in Vallée 2016).

The spiral arm model from Reid et al (2014 – their Fig.1; 2016 – their fig.1) has some differences with our model above.

(i) It fitted the few masers known, but it does not fit all the other tracers, like the CO 1-0 arm tangents as seen observationally in galactic longitudes (all in Table 3 to 10 in Vallée 2016).

(ii) The spiral arm model of Reid et al (2014; 2016) fits the masers in Galactic Quadrants I and II, but it does not fit the observations in the Galactic Quadrants III and IV (leaving out half of the Milky Way).

(iii) The publications with maser-derived pitch angle gave different pitch angle values along an arm. For instance, for the Scutum arm, the maser-derived published pitch angle values are: -7.0º (Reid 2012). -19.8º (Reid et al 2014); -19.2º (Krishnan et al 2015); -13º (Reid et al 2016 – their Fig.1); -18.7º (Xu et al 2018 – their table 4). Pitch angle differences near 10º are thus common in maser-derived values, although the published pitch angle error bar is near 1º. Similarly for the Sagittarius arm, the maser-derived published pitch angle values are: -11.2º (Sato et al 2010); -6.9º (Reid et al 2014); -19º (Krishnan et al 2017); -13.5º (Xu et al 2018 – their Table 4); -7.2 (Wu et al 2019). Some masers may be in an interarm or in a spur, wrongly assigned to the nearest arm, thus changing the arm pitch value.

(iv) Their maser-derived arm pitch angle in Galactic Quadrant I only differs notably from the global pitch angle found: employing a global arm fit over Galactic Quadrants I and IV (Fig. 1 and Table 1 in Vallée 2015) gave a pitch of -13º for the Sagittarius-Carina arm, a pitch of -13º for the Scutum-Crux-Centaurus arm (Table 2 in Vallée 2015), and a pitch of -13º for the Norma arm in both Galactic Quadrants I and IV (Table 1 in Vallée 2017a).

(v) The width of the maser lane may encompass both the masers located in the inner arm edge and the masers in spurs shooting out of the inner arm edge - thus artificially broadening the width of the maser lane (to 300 pc or so in Fig. 2 of Reid et al 2016).

(vi) The selected solar distance to the galactic center $R_{Sun}$ =8.34 kpc differs slightly from recently published values: 7.9±0.3 kpc (Qin et al 2018); 8.0±0.2 kpc (Vallée 2017d); 8.0±0.3 kpc (Camarillo et al 2018); 8.13±0.03 kpc (Abuter et al 2018); 8.18±0.03 kpc (Abuter et al 2019). These differences should not have a large impact on the present work, being within the distance uncertainties, and statistically the results on the offsets would be unchanged.

The Bayesian model used does employ an arm model from Reid et al (2014; 2016) and thus the assumptions in their arm models ($R_{Sun}$, pitch angle, arm width, rotation curve) may change with better data (see differences i to vi as noted above); also there is a possible systematic bias if the bayesian distance estimation is strongly influenced by the proximity of a chosen spiral arm. Some tests were made to look for such a bias (Section 2.3).

2.2 The physical offset

There is a physical offset between the shock lane (observed dust lane) and the predicted potential minimum (observed peak in low-density CO 1-0 gas) of a density wave. In our Milky Way galaxy, this offset is observed near 350 pc as an average for the 4 spiral arms (see Vallée 2016; Vallée 2017c).

Predictions vary a little, depending on many factors – thus Roberts (1975 – his Fig. 2) has an offset of 3.7% of the distance to the next arm, giving for a radius of 8 kpc and a 4-arm model a value of 465 pc. The offset shock-to-potential minimum is 3.2% for the model of Gittins & Clarke (2004 – their fig. 11), corresponding for a radius of 8 kpc and a 4-arm model to a value of 402 pc. In the model of Dobbs and Pringle (2010 - their Fig. 4a), the offset shock-to-potential minimum is near 5º (out of 180º) or about 2.8%, giving at a radius of 8kpc and a 4-arm model a value of 352 pc.

This offset (shock to potential minimum) defines very well an arm, in the observational sense.

2.3 Results and tests

Figure 1a shows the results for a top-view of the disk of the Milky Way. Here the 4-arm model has an arm pitch near -13.1°, and a Sun to Galactic Center distance of 8.0 kpc. Trigonometric maser distances and Bayesian filament distances are from Table 1.

In or near the Scutum-Crux-Centaurus arm, one sees 17 bone-like filaments, thus: 10 'Bones' (shown here as open squares), as well as 4 'mst' (open circles), and 3 'Herschel' (open triangles), all from Zucker et al (2018 – their table 1).

In or near the Scutum-Crux-Centaurus arm, a huge majority of the 'bone-like' filaments (14/17 = 82%) are located on the arm or on the outer side of the CO-fitted arm model (towards the anti-galactic center). Meanwhile, more than half of the radio masers (7/13 = 54%) are located on the arm or on the inner side of the CO-fitted arm model (towards the Galactic Center).

The separation of the 'bone-like' filaments, to the nearest point on the CO-fitted arm model, has a statistical mean of 130 pc (±36 pc) outward (away from the Galactic Center). Meanwhile, the mean separation of masers from the CO-fitted arm model has a mean of 78 pc (±72 pc) inward (towards the Galactic Center). Thus the separation of bone-like filaments from the masers is about 208 pc.

Tests. Changing some assumptions will not change much the results. Changing $R_{Sun}$ to 8.34 kpc for the 4-arm model, with the trigonometric maser and Bayesian filament data from Table 1, then the 4-arm model has to be enlarged, in order to keep matching the arm tangents in the CO 1-0 gas tracer, on both side of the Galactic Meridian (l=0°). The mean offset in distance between filaments and masers has not changed.

Scaling the current Bayesian distances for the filaments (employing $R_{Sun}$ =8.34 kpc) by using $R_{Sun}$ =8.00 kpc instead (to be the same as the 4-arm model with $R_{Sun}$ = 8.0 kpc), then one finds that the mean offset between filaments and masers has slightly increased outward (by roughly 50 pc, away from the masers).

Finally, to test whether there are significant biases being introduced by using the Bayesian distances, Figure 1b shows the results for the 4-arm model, with the masers' trigonometric distances from Table 1, and with the kinematical distances for the filaments (see Table 1). These kinematical distances were computed using equations 1 and 2 and Fig.1 in Roman-Duval et al (2009), with 8.0 kpc and 230 km/s as the Sun's distance to the Galactic Center and the orbital velocity of the Local Standard of Rest (same values as for the CO-fitted 4-arm model here). One can readily see in Figure 1b that a huge majority of filaments (12/17 = 71%) are on or above (outer side) of the CO-fitted arm, while a minority (5/17) is on the inner side. So, there is still a clear offset, between the parallax-derived maser distances and the kinematically-derived filament distances, and this offset is in the same direction (outwards of the masers) when compared to our previous Figure 1a. Hence this offset is not 'artificially introduced' by the Bayesian algorithm itself (as it is seen in both the kinematic distances alone, and in the Bayesian estimates).

Here the median distance offset increased for kinematic filaments (Fig.1b) as compared for Bayesian filaments (Fig. 1a). The kinematical distances come with a larger dispersion than the Bayesian distance estimates, since they assume that all the observed radial velocity is due to galactic rotation.

**Figure 2** shows the results for a longitude-velocity view. Here the 4-arm model has a circular velocity near 230 km/s, and an arm start at 2.2 kpc from the Galactic Center.

In or near the Scutum-Crux-Centaurus arm, one sees 17 bone-like filaments: 10 'bones' (open squares), 4 'mst' (open circles), and 3 'Herschel' (open triangles), from Zucker et al (2018 – their table 1).

In Galactic Quadrant I, in or near the Scutum-Crux-Centaurus arm, about one-half of the 'bone-like' filaments (7/13 = 54%) are located on the outside of the CO-fitted arm model 'loop' (towards lower radial velocities). Meanwhile, a clear majority of the radio masers (8/12 = 67%) are located on the inside of the CO-fitted arm model 'loop' (towards higher radial velocities).

In Galactic Quadrant I, the separation of the 'bone-like' filaments from the diffuse CO-fitted arm model has a mean of 5.2 km/s (± 2.8 km/s) outside of the loop (lower radial velocity). Meanwhile, the mean separation of masers from the CO-fitted arm model has a mean of 3.0 km/s (±2.6 km/s) inside of the loop (higher radial velocity). Thus the separation of bone-like filaments from the masers is about 8 km/s.

**Figure 3** shows a crosscut of the various arm tracers, as found earlier through the plot of the galactic longitude of the tangent from the Sun of each arm tracer, summarizing six arm segments: Scutum, Crux-Centaurus, Sagittarius, Carina, Norma, Perseus-origin (Vallée 2016; Vallée 2017c).

The mean position for the older stars (gray, near -70 pc) is from the data in table 10 in Vallée (2016) and table 2 in Vallée (2017c).

The mean position for the 'bone-like' filaments (gray, near –130 pc) is from the data for the Scutum-Crux-Centaurus arm (see above).

The arm width encompasses the region between the narrow space occupied by the masers (orange, at right), and the space occupied by the 'bone-like' filaments (gray, at left).

## 3. Predictive models for the origins and evolutions of filaments

Several theories have been published for the origins of large-scale filaments, yet none has gathered full support.

Perhaps there are many origins for these filaments. Some filaments could be born before star formation in the dust/shocked lane, and others could happen after the initial star-forming phase, being part of a 'second phase' of star formation stemming from older stars in a giant cluster going supernova with a huge supershell encompassing these older stars.

### 3.1 Galactic origin – a large-scale origin ?

Inter-arm formation, entering an arm ? Duarte-Cabral & Dobbs (2016) simulated the interstellar medium in a Milky-Way-type spiral galaxy, finding filaments forming in the interarm. Those filaments then become more well-defined as they reach the inner arm side or potential minimum (Duarte-Cabral & Dobbs 2016; 2017).

Duarte-Cabral & Dobbs (2017) followed the evolution of an interarm filament, entering in a spiral arm, finding that their morphology changes (becoming distorted and sub-structured) as it resides in the arm and leaves the spiral arm (from the outer edge of an arm, away from the Galactic Center direction).

This picture would be consistent with the observational result where the 'Nessie' filament was located near the middle of the Scutum arm in Fig. 3 of Goodman et al (2014), defined as the location of most optical stars, and near the middle of the radial velocity curve, defined for the CO gas from McClure-Griffiths & Dickey (2007).  However, with the recent 4-arm model (Vallée 2017a, b) shown here in Fig. 1a, along with the more accurate Bayesian distance to Nessie (Zucker et al 2018),  it is now found that the CO-arm to Nessie displacement is 200 pc outward (see Table 1 and Fig.1) – farther away from the masers. While each filament distance in the Bayesian method has an estimated error near 300 pc (Fig. 4 in Zucker et al 2018), the fact that most filaments in Figure 1a are outside the CO 1-0 arm model implies a statistical mean displacement of bone-like filaments to the CO-fitted arm, with a smaller statistical uncertainty (± 36 pc – see Section 2).  It is not clear  how this picture fits with the numerical models from Duarte-Cabral & Dobbs (2016, 2017), as those studies did not explore the type of structures exiting the spiral arm. The type of filamentary structures detected in those models are, in fact, most likely better matched to the 'gmf' type of filaments, which shows a more random distribution in the Galaxy.

Figure 4 here shows the location of the 'gmf' filaments near the Scutum arm (from Table 1 in Zucker et al 2018), with Bayesian distance estimates, plotted on the same map used in Fig. 1. It is clear that the 'gmf' filaments are not distributed in the same way (being broader, almost horizontal) than the 'bone-like' distribution (narrower, closer to the spiral arm).

In-arm formation, exiting an arm? Smith el al (2014a) performed similar MW-type galaxy simulations, showing the 'dark molecular gas' or 'CO-dark regions' or gas clouds in spiral arms, and exiting an arm as dense filamentary 'spurs' being sheared off an arm by galactic rotation (their Fig. 7). Smith et al (2014b) found that a filament could be made up of a network of shorter 'ribbons' or 'fibers' or 'sub-filaments'.

Some filaments could be located in the mid arm or spurs as the gas gets out of the potential minimum, and starts getting stretched by the galactic shear (Smith et al 2014a). Note that Smith et al (2014a) did not include self-gravity, star-formation, or feedback, thus it is hard to infer what would happen to these long filaments with the inclusion of more physics.

3.2 Local origin  –  a small-scale origin ?

Some filaments could be part of a 'second phase' of star formation originating from older stars in a giant cluster with a supershell. This second phase could be downward from the dust/shock lane.

Some filaments, observed in great details, appear to be in pressure equilibrium (Vallée 2007 – his fig. 2; Vallée & Fiege 2006 – their section 5.5) near an ensemble of old stars, located past the middle of a spiral arm (near the outer arm edge, past the star forming region). They could thus be 'local' (2$^{nd}$ phase, now) while also originally be 'galactic' (1$^{st}$ phase, long ago).

For example, radio maps showed a molecular filament, linking protostars in DR21(OH) to those in DR21, ERO02, and ERO03, extending about 22 pc long and 3 pc wide (fig. 1 in Vallée & Fiege 2006), at galactic longitude 81.7° and latitude +0.6°, and at a velocity distance near 2.5 kpc (not far from the Sagittarius arm, between the Sagittarius and Perseus arms). Its median column density is near $30 \times 10^{21}$ cm$^{-2}$. This elongated filament (length / width near 7) is also near the Galactic plane (about 26 pc). This molecular filament was studied intensively in spectroscopy (rotation, turbulence) and polarimetry (orientation, magnetic field), using the CO, $^{13}$CO and C$^{18}$O J=3-2 emission lines. Vallée & Fiege (2006) found that this filament is not spinning (no rotation across its axis) nor tumbling (no rotation along the axis). The turbulence (turbulent line width) is the same along and across the filament; this filament is not expanding. The large-scale magnetic field is essentially perpendicular to the long axis of the filament, with a strength near 200 µgauss. Its turbulent energy is about equal to its magnetic energy, each near 400x10$^{34}$ ergs. This filament's internal pressure balances the external pressure of its surroundings (Section 5.5 in Vallée & Fiege 2006). It was argued there that this filament could have formed by compression inside an expanding shell (due to wind-blown bubble around an OB association or due a supernova remnant) and the filament drifted into the interior cavity of the shell (due to the outward expansion of the shell). This compression model (inside an expanding shell, diameter near 440 pc) would involve older stars, and would involve a location past the star-forming region in a spiral arm (where older stars are located, as well as a new generation of O-B associations). Schneider et al (2010) and Hennemann et al (2012) studied the smaller scale, and favored a very slow infalling model (0.6 km/s) in order to build the inside of the 22-pc filament and its numerous starforming clumps. This highly visible DR21 filament is on the outskirts of the larger (7° or 300pc) Cygnus-X complex of current starformation.

A similar molecular filament, linking OMC-1 to OMC-2 and OMC-3 and OMC-4 and others covering 4°, is about 31 pc long and 1 pc wide (fig. 1 in Vallée & Fiege 2007), at galactic longitude 209.0° and latitude -19.4°, and at a distance near 0.5 kpc (close to the Sagittarius-Carina arm, between the Sagittarius and Perseus arms). Its median column density is near 35x10$^{21}$ cm$^{-2}$. This elongated filament (length/width =31) is near the Galactic plane (about 160 pc). That filament could also have formed by compression in an expanding shell (a wind-blown bubble around an OB association, with a diameter near 120 pc) and would later drift in the cavity of the shell (outward shell expansion). Indeed, a pressure balance (internal gas versus external gas) was also found for the OMC-1 filament (Fig. 2 in Vallée 2007). This compression model involves older stars. Numerous others studied the smaller scale of the OMC, notably OMC-3 (Takahashi et al 2013), OMC-2 (Sadavoy et al (2016), OMC-1 (Texeira et al 2016). This 31-pc filament may be related to the 700-pc expanding Gould's Belt near the Sun.

4. Discussion
4.1 Predictive models

The predictive models proposed above for the origin and evolution of a 'bone-like' filament can be compared to our statistical results. These observed high-density 'bone-like' structures may form a sub-sample of larger low-density elongated clouds. Current numerical theories may more easily detect the latter, not the former. Numerical models are probably more capable of probing structures on the scales of gmf, but could have trouble seeing things like bone-like filaments (too small, too dense).

It may be worth mentioning that the 'gmf' may have a spatial distribution differing substantially than that for the 'bone-like' filaments – see Fig. 4 and Section 3.1.

For the Scutum-Crux-Centaurus arm, the more precise distance estimates afforded by the Bayesian method (filaments) and the trigonometric method (masers) indicate a separation of these two (Fig. 1a). The masers appear more on the inner side of the spiral arm (towards the Galactic Center), while the 'bone-like' filaments appear more on the outer side of the arm (away from the Galactic Center). This offset can also be seen in the poorer kinematic distances of the filaments (Fig. 1b).

This separation of masers versus bone-like filaments is predicted, when 'bone-like' filaments are created in shells and cavities around stars (Vallée & Fiege 2006), as the distribution of old stars appear to peak towards the outer arm edge (away from the Galactic Center).

4.2 Observational models

The observational arm model employed here was explained elsewhere, as a fit to the arm tangents observed in the diffuse CO 1-0 tracer (Vallée 2016; Vallée 2017a; Vallée 2017b), thus including a few thousands of individual cells of CO-emitting gas. Its fitted arm pitch angle for Scutum is close to -13º (Vallée 2015 – his table 2). The width of the maser lane is close to 100 pc or so (see Fig. 3 here).

Zucker et al (2018 – their Section 3.7) strongly emphasized that the arm traces in the Reid et al (2016) model are not necessarily correct or complete, and that their results on filaments only hold in the context of the arm model of Reid et al (2016), and could change with the adoption of different spiral arm models.

There should be some spatial differences between these two arm models (masers versus diffuse CO), as the dust lane and maser lane are separated from the diffuse CO arm center by about 300 pc or so (see Fig. 3 here). Some comparisons with several other arm models are given in Vallée (2017b – his table 3).

5. Conclusion

The 'bone-like' filaments and masers, with a Bayesian or trigonometric distance estimates, near the Scutum-Crux-Centaurus arm (309º < l < 033º), are employed to search for a relationship with a recent spiral arm model, in both geometric (Fig.1a) and velocimetric (Fig.2) spaces.

The 17 'bone-like' filaments in Zucker (2018) were plotted in a recent top-view and longitude-velocity 4-arm model, to find where these 'bone-like' filaments were located close to the Scutum-Crux-Centaurus spiral arm.

The results imply that many 'bone-like' filaments are found on the outer side of the Scutum-Crux-Centaurus arm (away from the Galactic Center), while the masers are preferentially found on the inner arm side. Here we find a separation between the maser lane and the 'bone-like' filaments of about 208 pc (Fig.1a). This separation is also seen using the poorer kinematic distances for filaments (Fig. 1b). Their likely position in a crosscut of a spiral arm is shown in Fig.3.

The 4$^{th}$ type (gmf), with more massive filaments and without 'bone-like' properties, have been plotted (Fig. 4), showing a largely random distribution, despite using Bayesian distance estimates. They may have a different origin or/and formation or/and evolution than the simpler bone-like filaments, perhaps more in line with the idea of inter-arm formation.

Acknowledgements. The figure production made use of the PGPLOT software at NRC Canada in Victoria. I thank an anonymous referee for useful, careful, and historical suggestions.

ORCID ID   Jacques P Vallée    https://orcid.org/0000-0002-4833-4160

References
Abreu-Vicente, J., Ragan, S., Kainulainen, J., et al 2016, A&A, 590, A131 ((20pp).
Abuter, R., Amorim, A., Anugu, N., and others. 2018, A&A, 615, L15 (10pp).
Abuter, R., Amorim, A., Baubock, M., and others. 2019, A&A, 625, L10 (10pp).
Camarillo, T., Mathur, V., Mitchell, T., Ratra, B. 2018, PASP, 130, 024101 (10pp).
Duarte-Cabral, A., Dobbs, C.L. 2016, MNRAS, 458, 3667-3683.
Duarte-Cabral, A., Dobbs, C.L. 2017, MNRAS, 470, 4261-4273.
Dobbs,C., Pringle, J. 2010, MNRAS, 409, 396-404.
Gittins, D., Clarke, C. 2004. MNRAS, 349, 909-921.
Goodman, A.A., Alves, J., Beaumont, C.N., et al. 2014, ApJ, 797, 53 (13pp).
Hennemann, M., Motte, F., Schneider, N., et al 2012, A&A, 543, L3 (7pp).
Krishnan, V., Ellingsen, S.P., Reid, M.J., et al 2017, MNRAS, 465, 1095 (21pp).
Krishnan, V., Ellingsen, S., Reid, M., et al 2015, ApJ, 805, 129 (12pp).
Matter, M., Kauffmann, J., Csengeri, T., Urquhart, J.S., et al. 2018, A&A, 619, 166 (24pp).
McClure-Griffiths, N.M., Dickey, J.M. 2007, ApJ, 671, 427-438.
Qin, W., Nataf, D., Zakamska, N., et al 2018, ApJ, 865, 47 (13pp).
Ragan, S., Henning, T., Tackenberg, J., et al 2014, A&A, 568, A73 (22pp).
Reid, M.J. Proc. 2012, IAU Symp., 289, 188-193.
Reid, M.J., Dame, T.M., Menten, K.M., Brunthaler, A. 2016, ApJ, 823, 77 (11pp).
Reid, M.J., Menten, K.M., Brunthaler, A., Zheng, X.W., et al. 2014, ApJ, 783, 130 (14pp).
Reid, M.J., Menten, K.M., Zheng, X.W., Brunthaler, A., and 10 others. 2009, ApJ, 700, 137-148.
Roberts, W.W. 1975, Vistas in Astron., 19, 91-109.
Roman-Duval, J., Jackson, J., Heyer, M. and others. 2009, ApJ, 699,1153 (18pp).
Sadavoy, S., Stutz, A., Schnee, S., et al 2016, A&A, 588, 30 (11pp).
Sato, M., Hirota, T., Reid, M., Honma, M., et al 2010, PASJ, 62, 287 (13pp).
Schneider, N., Csengeri, T., Bontemps, S., et al 2010, A&A, 520, A49 (21pp).

Smith, R.J., Glover, S.C., Clark, P.C., Klessen, R.S. Springel, V. 2014a, MNRAS, 441, 1628-1645.
Smith, R.J., Glover, S.C., Klessen, R.S. 2014b, MNRAS, 445, 2900-2917.
Takahashi, S., Ho, P., Texeira, P., Zapata, L., Su, Y. 2013, ApJ, 763, 57 (13pp).
Texeira, P., Takahashi, S., Zapata, L., Ho, P. 2016, A&A, 587, 47 (10pp).
Vallée, J.P. 2007, ApJ, 134, 511-515.
Vallée, J.P. 2014, Astro. J., 148, 5 (pp).
Vallée, J.P. 2015, MNRAS, 450, 4277-4284.
Vallée, J.P. 2016, ApJ, 821, 53 (12pp).
Vallée, J.P. 2017a, Astrophys. Space Sci., 362, 173 (5 pp).
Vallée, J.P. 2017b, New Astron. Rev., 79, 49-58.
Vallée, J.P. 2017c, Astron. Rev., 13, 113-146.
Vallée, J.P. 2017d, Ap Sp Sci., 362, 79 (6 pp).
Vallée, J.P., Fiege, J.D. 2006, ApJ, 636, 332-347.
Vallée, J.P., Fiege, J.D. 2007, AJ, 133, 1012-1026.
Wang, K., Testi, L., Ginsburg, A., et al 2015, MNRAS, 450, 4043-4049.
Wang, K., Testi, L., Burkert, A., et al 2016, Ap J Suppl, 226, 9 (17pp).
Wu, Y.W., Reid, M.J., Sakai, N., Dame, T., et al 2019, ApJ, 874, 94 (13pp).
Xu, Y., Hou, L., Wu, Y. 2018, Res. Astron. Astrophys., 18, 146 (20pp).
Zhang, M., Kainulainen, J., Mattern, M., Fang, M., Henning, Th. 2019, A&A, 622, 52 (57pp).
Zucker, C., Battersby, C., Goodman, A. 2015, ApJ, 815, 23 (25pp).
Zucker, C., Battersby, C., Goodman, A. 2018, ApJ, 864, 153 (34pp).

Table 1. Sources near the Scutum arm ($309° < l < 033°$), with a measured Bayesian or trigonometric distance.

| Name | Gal. Long. (o) | Gal. Lat. (o) | Distance (kpc) | Syst. $V_{lsr}$ (km/s) | Reference |
|---|---|---|---|---|---|
| Filaments (Note 1): | | | | | |
| Fil 7 – bone | 04.1 | -0.0 | 2.9/2.6 | +8 | Zucker et al (2018) |
| Fil 6 – bone | 11.1 | -0.1 | 3.0/3.4 | +31 | Zucker et al (2018) |
| F10 – mst | 12.9 | -0.2 | 3.0/3.4 | +35 | Zucker et al (2018) |
| F15 – mst | 14.2 | -0.2 | 3.2/3.5 | +40 | Zucker et al (2018) |
| F14 – mst | 14.7 | -0.2 | 3.1/3.4 | +39 | Zucker et al (2018) |
| Fil 5 – bone | 18.9 | -0.1 | 3.4/3.4 | +46 | Zucker et al (2018) |
| Fil 3 – bone | 24.9 | -0.2 | 3.5/3.1 | +47 | Zucker et al (2018) |
| Fil 2 – bone | 25.2 | -0.4 | 3.7/3.5 | +57 | Zucker et al (2018) |
| F28 – mst | 25.3 | -0.2 | 3.7/3.8 | +63 | Zucker et al (2018) |
| G26 – herschel | 26.4 | +0.8 | 3.0/3.0 | +48 | Zucker et al (2018) |
| Fil 1 - bone | 26.9 | -0.3 | 4.1/3.9 | +68 | Zucker et al (2018) |
| G28 – herschel | 28.7 | -0.3 | 4.7/4.8 | +88 | Zucker et al (2018) |
| G29 – herschel | 29.2 | -0.3 | 5.0/5.1 | +94 | Zucker et al (2018) |
| Fil 10 – bone | 332.2 | -0.0 | 3.2/3.1 | -49 | Zucker et al (2018) |
| Fil 9 – bone | 335.3 | -0.3 | 2.9/2.8 | -42 | Zucker et al (2018) |
| Nessie- bone | 338.5 | -0.4 | 2.8/2.8 | -38 | Zucker et al (2018) |
| Fil 8 – bone | 357.6 | -0.3 | 2.8/2.4 | +4 | Zucker et al (2018) |
| Masers (Note 2): | | | | | |
| G005.88-00.39 | 005.9 | -0.4 | 2.99 ±0.17 | +9 | Reid et al (2014) |
| G011.91-00.61 | 011.9 | -0.6 | 3.37 ±0.32 | +37 | Reid et al (2014) |
| G012.80-00.20 | 012.8 | -0.2 | 2.92 ±0.29 | +34 | Reid et al (2014) |
| G013.87+00.28 | 013.9 | +0.3 | 3.94 ±0.34 | +48 | Reid et al (2014) |
| G016.58-00.05 | 016.6 | -0.1 | 3.58 ±0.33 | +60 | Reid et al (2014) |
| G023.00-00.4 | 023.0 | -0.4 | 4.59 ±0.33 | +81 | Reid et al (2009) |
| G028.86+00.06 | 028.9 | +0.1 | 7.41 ±1.13 | +100 | Reid et al (2014) |
| G029.86-00.04 | 029.9 | -0.0 | 6.21 ±0.77 | +100 | Reid et al (2014) |
| G029.95-00.01 | 030.0 | -0.0 | 5.26 ±0.58 | +98 | Reid et al (2014) |
| G031.28+00.06 | 031.3 | +0.1 | 4.27 ±0.73 | +109 | Reid et al (2014) |
| G031.58+00.07 | 031.6 | +0.1 | 4.90 ±0.76 | +96 | Reid et al (2014) |
| G032.04+00.05 | 032.0 | +0.1 | 5.18 ±0.21 | +97 | Reid et al (2014) |
| G348.70-01.04 | 348.7 | -1.0 | 3.38 ±0.27 | -7 | Reid et al (2014) |

Note 1: Names of all filaments identified as located in the 'Scutum' arm in Zucker et al (2018 – their column 6), with the type: 'bone', 'mst', or 'herschel' (their column 5). Distances at left are Bayesian; distances at right are kinematical (see text).

Note 2: Taking all trigonometric maser sources, identified as located in the 'Scutum' arm in Reid et al (2014) and Reid et al (2009), excluding 3 sources with a nearby solar distance < 2.6 kpc (G12.68; G12.88; G12.90), and excluding 2 sources with a large distance error > 2.0 kpc (G25.70; G27.36). All distances are trigonometric; the published parallax (p, in mas) was converted to a distance (D, in kpc) through the equation D = 1/p.

Figure captions.

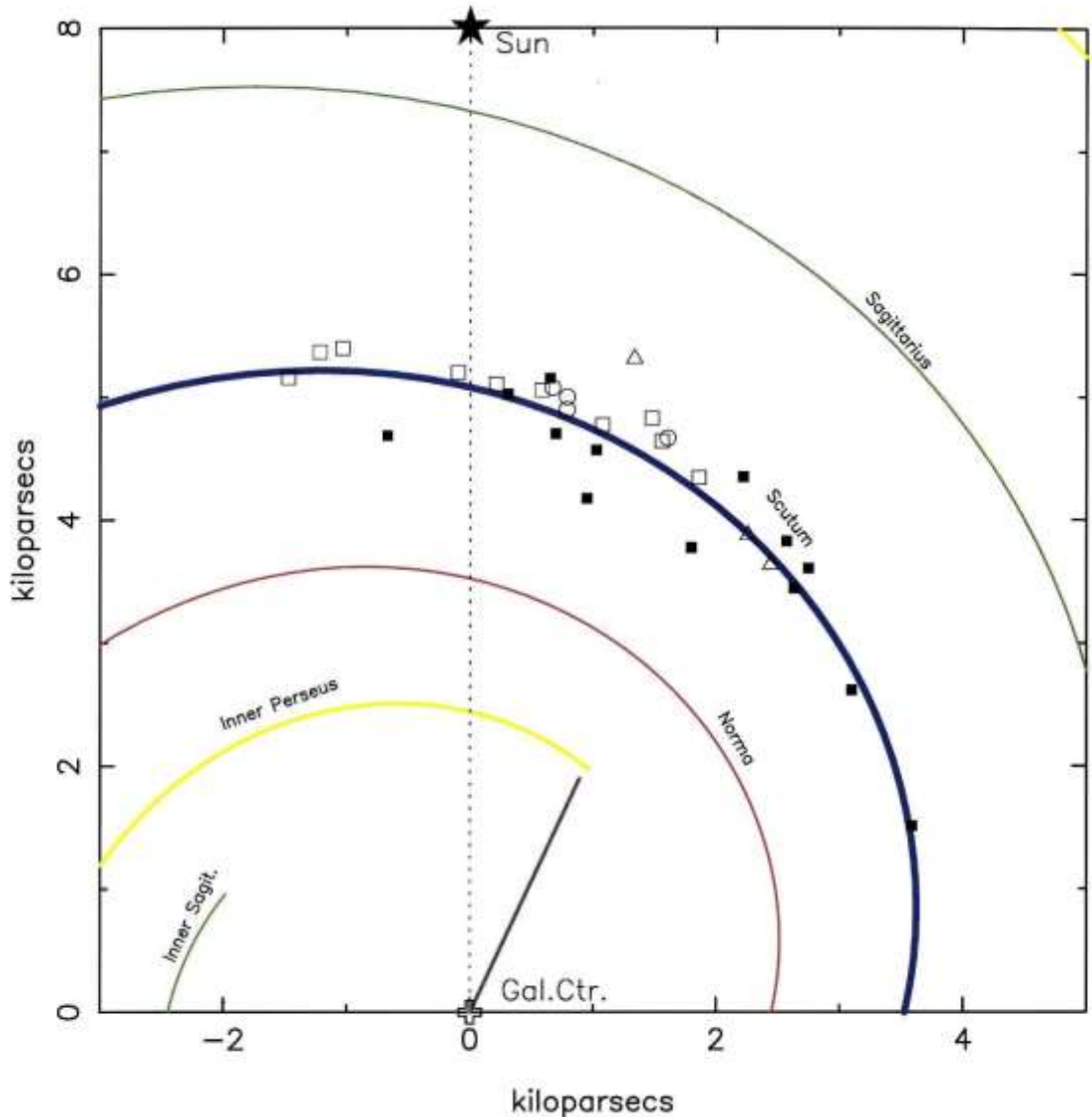

Figure 1a.  Top view of the Milky Way disk, with the Sun at top and the Galactic Center at bottom. Galactic quadrants I (at right) and IV (at left) are shown.

A recent spiral model of 4 spiral arms is shown (Vallée 2017a; 2017b), as fitted to the observed diffuse CO arm tangents (Vallée 2016). The Scutum-Crux-Centaurus arm is shown in blue. Here the distance of the Sun to the Galactic Center is set at 8.0 kpc.

In or near the Scutum-Crux-Centaurus arm, Zucker et al (2018 – their table 1) identified 10 'Bones' (open squares), 4 'MST' (open circles), and 3 'Herschel' (open triangles). Bayesian distances were used for the filaments. While the individual error bar for each filament is about 0.3 kpc, the statistical mean done on the ensemble of filaments is better by a factor of 3 or more; data in Table 1. In or near the Scutum-Crux-Centaurus arm, Reid et al (2014 – their table 1) identified 14 masers, outside a solar distance of 2.6kpc (shown here as filled squares). Trigonometric distances were used for the masers - data in Table 1.

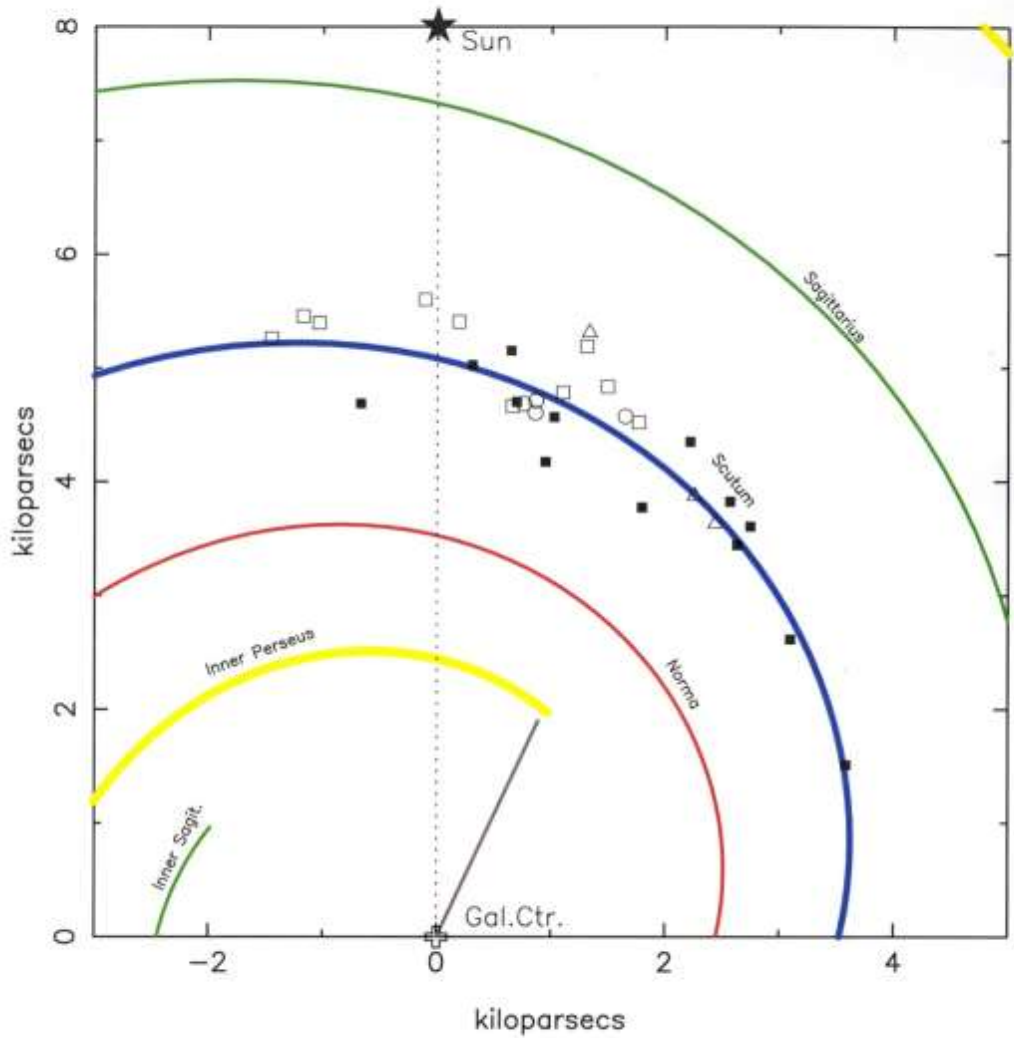

Fig.1b.    Here, kinematic distances for the filaments were used – data in Table 1.  In or near the Scutum-Crux-Centaurus arm, Zucker et al (2018 – their table 1) identified 10 'Bones' (open squares), 4 'MST' (open circles), and 3 'Herschel' (open triangles)..   In or near the Scutum-Crux-Centaurus arm, Reid et al (2014 – their table 1) identified 14 masers, outside a solar distance of 2.6kpc (shown here as filled squares); their trigonometric distances were used - data in Table 1.

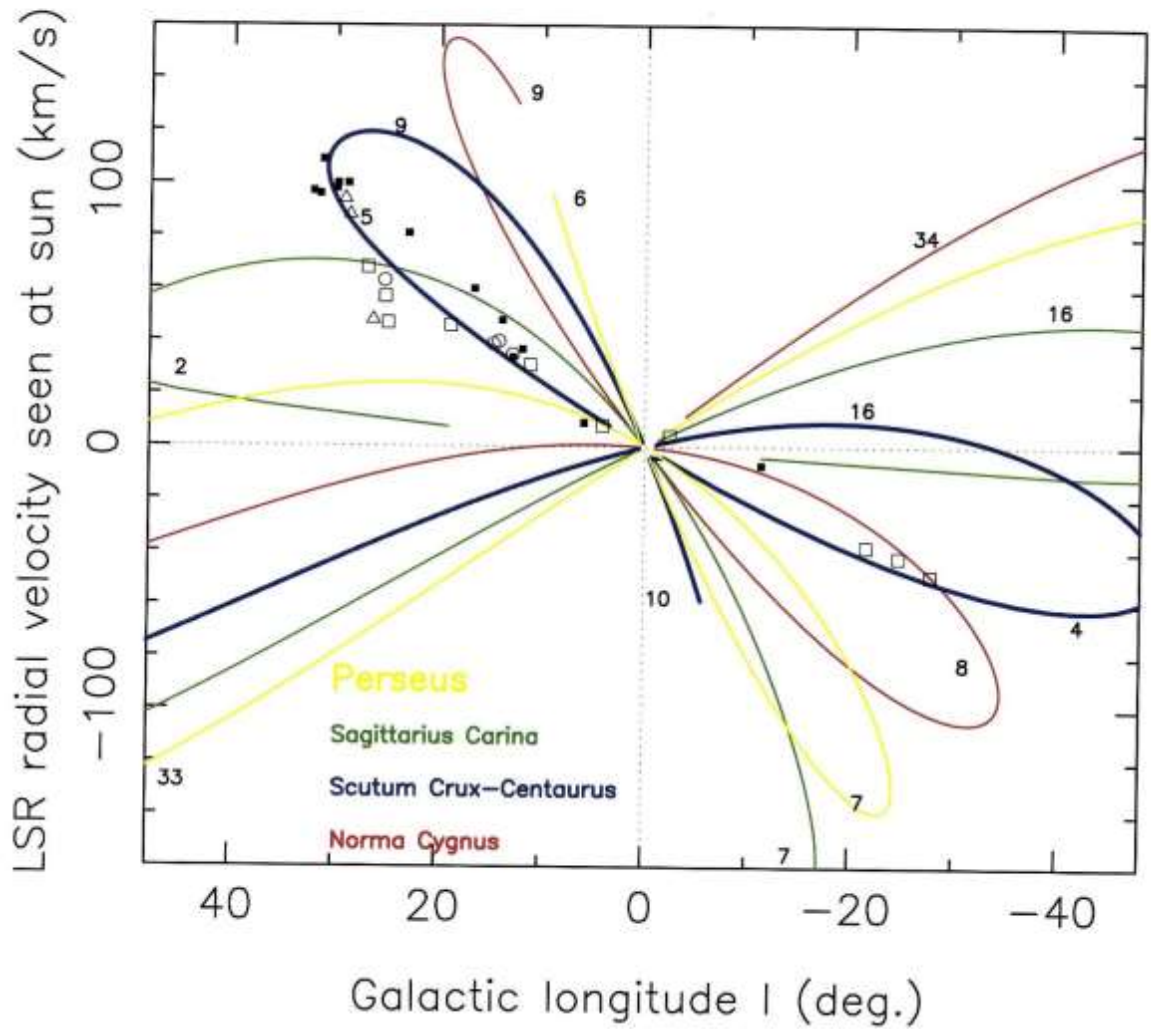

Figure 2. A diagram of galactic longitudes versus radial velocity for observed objects..Here Galactic quadrants I (at left) and IV (at right) are shown. Our recent spiral velocimetric model of 4 spiral arms is shown (Vallée 2017a, 2017b), as fitted to the observed diffuse CO arm tangents (Vallée 2016). Numbers on the arm model indicate a rough distance of that point from the Sun. The Scutum-Crux-Centaurus arm is shown in blue.

Zucker et al (2018 – their table 1) identified 10 'Bones' (open squares), as well as 4 'MST' (open circles), and 3 'Herschel' (open triangles) – data from Table 1.

In or near the Scutum-Crux-Centaurus arm, Reid et al (2014 – their table 1) identified 14 masers, outside a solar distance of 2.6kpc (shown here as filled squares) – data from Table 1.

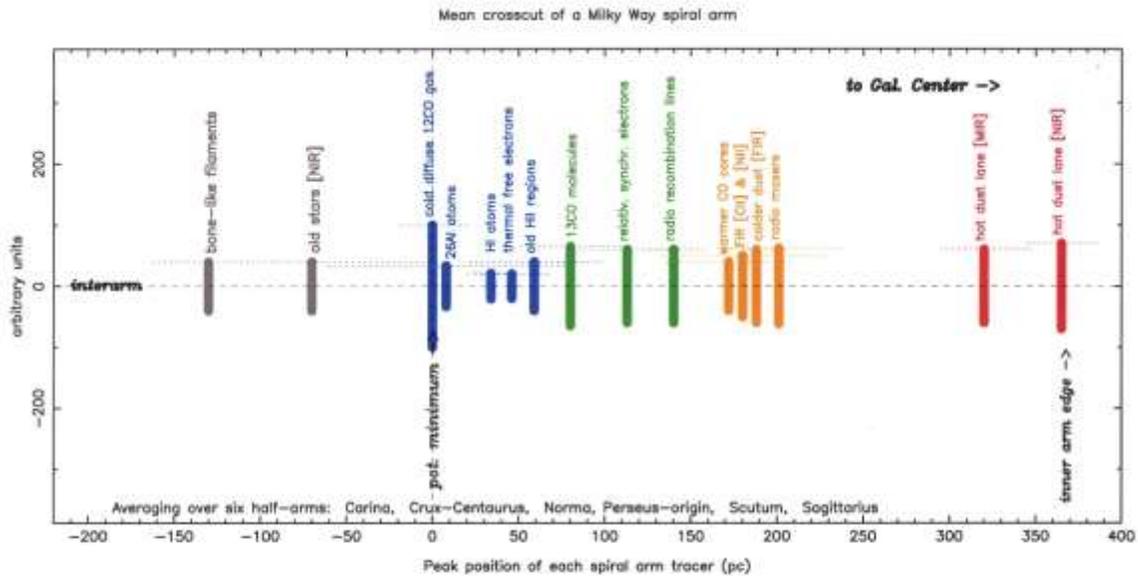

Figure 3. A crosscut of a typical spiral arm in the Milky Way disk. The inner arm edge is to the right; the direction to the Galactic Center is to the right. Six arm segments were averaged. For each arm segment, observations for each observed tracer were averaged over galactic longitudes (Vallée 2016). The y-axis has arbitrary units. Error bars for the statistical means are shown horizontally with dots. The interarm region is shown at far left.

One sees the inner arm edge at right (red hot dust, orange radio masers and FIR [CII] & [NII] lines), the middle of the star-forming region (green radio synchrotron electrons and recombination lines, blue diffuse CO 1-0 gas and old HII regions), and a region further out at left (gray old stars and bone-like filaments).

In the density-wave model, the shocked dust lane is at right (observed around x=350 pc). The 'potential minimum' in density-wave amplitude is near the peak intensity of the cold diffuse CO 1-0 gas as observed with a broad telescope beamwidth (at x= 0). Adapted from Vallée (2017c).

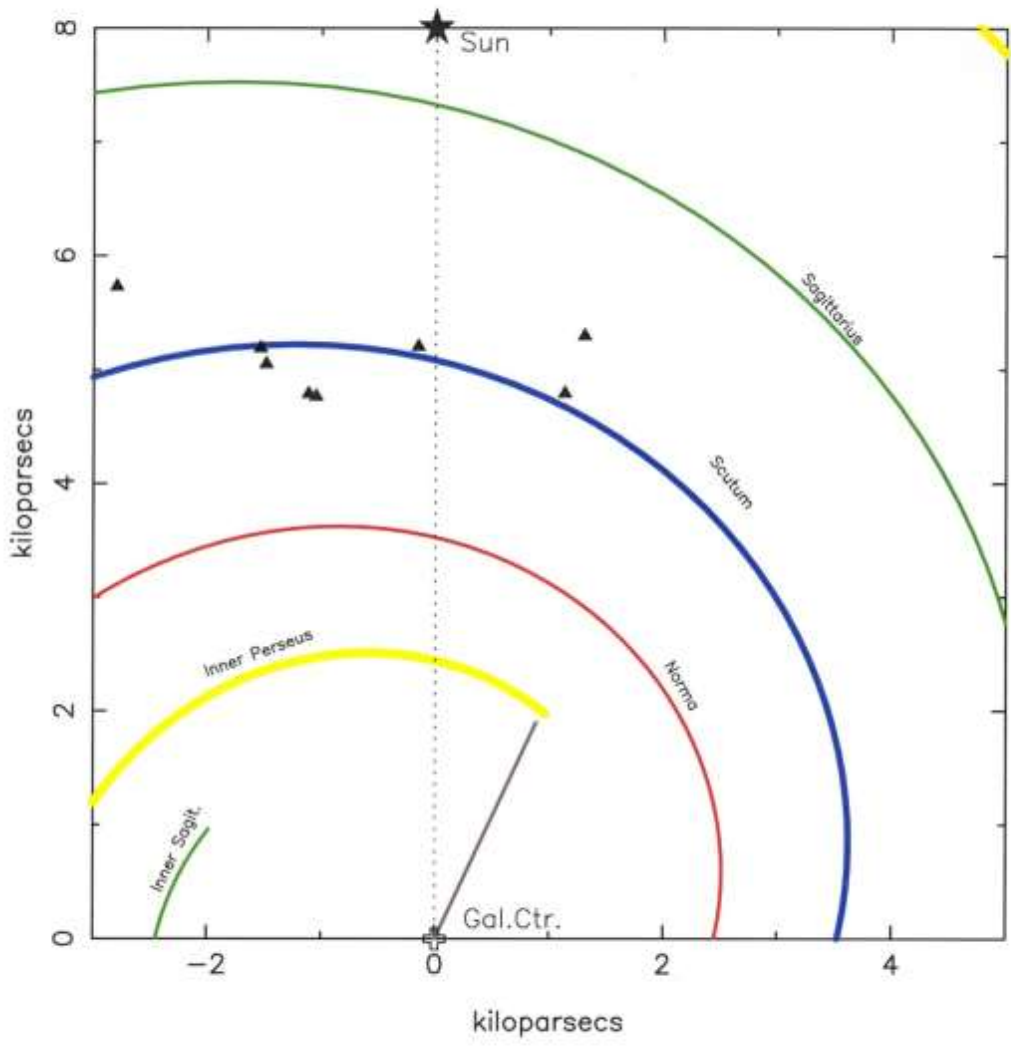

Fig. 4. In or near the Scutum-Crux-Centaurus arm, Zucker et al (2018 – their table 1) identified 8 'gmf' (black triangles). Their locations seem to be wider than the bone-like filaments, more horizontal, not apparently aligned to the CO-fitted Scutum arm model. The 4-arm model is that of Figure 1.